\documentclass[prb, preprint]{revtex4}
\usepackage{amsfonts}
\usepackage{amsm ath}
\usepackage{amssymb}
\usepackage{graphicx}
\usepackage{epstopdf}
\usepackage{color}
\usepackage{bbm}
\usepackage[T1]{fontenc}
\usepackage{bbold}

\input epsf.sty 

\bibpunct{[}{]}{;}{n}{}{}

\begin{document}

\title{Tunable topological semi-metallic phases in Kondo lattice systems}
\author{Yen-Wen Lu$^{1}$, Po-Hao Chou$^{1}$, Chung-Hou Chung$^{3}$, and Chung-Yu Mou $^{1,2,4}$}
\affiliation{$^{1}$Center for Quantum Technology and Department of Physics, National Tsing Hua University, Hsinchu 30043,
Taiwan, 300, R.O.C.}
\affiliation{$^{2}$Institute of Physics, Academia Sinica, Nankang 115, Taiwan, Republic of China}
\affiliation{$^{3}$Electrophysics Department, National Chiao-Tung University, HsinChu, Taiwan, R.O.C.}
\affiliation{$^{4}$Physics Division, National Center for Theoretical Sciences, Hsinchu 30013, Taiwan, Republic of China}
\begin{abstract}
We exploit topological semi-metallic phases resulting from the Kondo screening in  Anderson lattice models. 
It is shown that by including spin-orbit interactions both in the bulk electrons and in the hybridization between the conduction electrons and electrons in $f$ orbit, all types of topological semi-metallic phases can be realized in Anderson lattice models.
Specifically, upon either time-reversal symmetry broken or inversion symmetry broken, we find that either Weyl semi-metallic phase, Dirac semi-metallic phase or nodal-ring semi-metallic phases always emerge between insulating phases and can be accessed by tuning either temperature or spin-orbit interaction. For Anderson lattice models with general 3D spin-orbit hybridization between the conduction electrons and electrons in $f$ orbit, we find that Weyl nodal-ring semi-metallic phases emerges between strong and weak topological insulating phases. Furthermore, in the presence of an exchange field, Weyl semi-metallic phases forms after two Weyl points of charge $\pm1$ split off from a Dirac point at time-reversal momenta. 
On the other hand, when the spin-orbit interaction is included in the conduction electron, we find that upon the rotation symmetry being broken with anisotropic hopping amplitudes, Weyl semi-metallic phase emerges with double Weyl node of charges of $\pm2$. Furthermore, the Weyl semi-metallic phases with charges of $\pm2$ can be tuned into Weyl semi-metallic phases with charges of $\pm1$ through the inclusion of the Rashba spin-orbit interaction. Our analyses indicate that Anderson lattices with appropriate spin-orbit interactions provide a platform for realizing all types of topological semi-metallic phases.
\end{abstract}
\maketitle
\section{Introduction}
Topological semimetals have recently attracted a lot of attention in condensed matters physics due to their bulk massless electronic structures 
and the presence of surface states in specific surfaces\cite{Weyl1, Weyl2, Weyl3, mou1}.  Starting from graphene discovered in 
2004\cite{Geim1,Geim2}, in which massless 2D Dirac fermions are realized, several materials that realize 3D version of semimetals: Dirac semimetals (e.g., Na$_3$Bi\cite{Na3Bi}, Cd$_3$As$_2$\cite{Cd2As3_1,Cd2As3_2}) and Weyl semimetals (e.g., TaAs\cite{TaAs, TaAs1, TaAs2}) are found subsequently. 
More recently,  massless points that form nodal lines are further found in PbTaSe$_2$\cite{PbTaSe2} and ZrSiS\cite{ZrSiS}.
In these materials, the conduction and valence bands cross at either discrete points (Dirac or Weyl semimetal\cite{Nagaosa}) or at lines (nodal line \cite{nodal_line}or ring semimetals) in the Brillouin zone (BZ). The crossing points are protected by symmetries\cite{Fang} and are responsible for a number of novel transport properties (such as  the anomalous Hall effect and the chiral magnetic effect)  in these materials. While these topological semimetals are usually formed by symmetries with fixed electronic structures, the corresponding semi-metallic phase is the critical phase that controls
phase transitions between two insulating phases with different topological properties. In particular, this implies that semi-metallic phases can be 
accessed through phase transitions. In the case of discrete massless points, it is known that  the mass of the Dirac point controls the transition between the topological trivial and the topological nontrivial phases\cite{Sato}.  Right at the point when the mass vanishes, the material is a Dirac semimetal which is  at a quantum critical point between the hole Fermi liquid and the electron Fermi liquid\cite{Sheehy}. Recently, it is shown that the Kondo screening in  Anderson lattices provides a platform to access the Dirac semimetallic critical point\cite{Mou}.  The semi-metallic critical point is also shown to be realized in a Kondo-Heisenberg Hamiltonian in integer filling of electrons\cite{filling}.  In particular, the electronic structure due to
the Kondo screening depends on temperature\cite{Coleman, DMFT1, DMFT2} so that the Dirac semi-metallic phase can be also accessed by tuning temperatures\cite{Mou}. 
Indeed, there has been several reports indicating that tuning temperatures or spin-orbit coupling strength to assess semi-metallic phases are feasible experimentally\cite{Dzsaber, Lai, 2D_temperature, HgTe}. Motivated by these observations, it is therefore appealing to explore possible semi-metallic phases that can be accessed by the  Kondo screening in  Anderson lattices. 

In this work, we explore topological semi-metallic phases resulting from the Kondo screening in simple cubic Anderson lattices. 
It is shown that by including appropriate spin-orbit interactions, all types of topological semi-metallic phases can be realized in simple cubic Anderson lattices.
Specifically, we shall show that upon either time-reversal symmetry broken or inversion symmetry broken,  either Weyl semi-metallic phase, Dirac semi-metallic phase or nodal-line semi-metallic phases always emerge between insulating phases with different topological properties.
Furthermore, these topological semi-metallic phases can be accessed by tuning either temperature or spin-orbit interaction. 
Our results reveal the unusual interplay between the topology of the
electronic structures and the Kondo screening in the strongly correlated Anderson lattices and pave a way for systematically engineering 
topological semimetals based on Kondo lattice systems.  

The rest of the paper is organized as follows. In Sec. II, the generic Anderson lattice model on a simple cubic lattice  with spin-orbit interactions
is introduced. By using the slave-boson method, the mean-field Hamiltonian is
constructed. In Sec. III,  we examine the Anderson lattice model without time-reversal symmetry. The resulting Weyl semi-metallic phase is inversion symmetric with Weyl nodes being split off from a Dirac point at time-reversal momenta. In Sec. IV,  we show that the Weyl nodal-ring semi-metallic phase generally emerges when the inversion symmetry is broken in the Anderson lattice model with general hybridization between the conduction electron and $f$ electron.  Sec V is denoted to investigate the Anderson lattice without inversion symmetry through the bulk spin-orbit interaction. It is shown that depending on the nature of spin-orbit interaction in hybridization, the emergent Weyl semi-metallic phase can host Weyl nodes with monopole charges being $\pm$ or double Weyl nodes with charges being $\pm 2$. Finally, in Sec. VI, we conclude and discuss possible effects due to fluctuations that go beyond the mean-field theory.
\section{Theoretical Model}
We start by considering the generic Anderson lattice model that includes spin-orbit interactions on a simple cubic lattice. The
model that describes effects of Kondo screening can be generally described by the following Hamiltonian
\begin{eqnarray} \label{Eq1}
H &=& \sum_{\mathbf{k}\sigma} (\xi_{\mathbf{k}} c_{\mathbf{k} \sigma}^\dag c_{\mathbf{k} \sigma} + \xi^d_{\mathbf{k}} d_{\mathbf{k} \sigma}^\dag d_{\mathbf{k} \sigma}) +H_{SO}\nonumber \\ 
&+& \sum_{\mathbf{k} \sigma \sigma' }  (V^{\sigma \sigma' }_{\mathbf{k}} c_{\mathbf{k} \sigma}^\dag d_{\mathbf{k} \sigma'}+ h.c.)  + U\sum_i n_{i\uparrow}^dn_{i\downarrow}^d. \label{ALH}
\end{eqnarray}
Here $c_{\mathbf{k} \sigma}^\dag$ and $d_{\mathbf{k} \sigma}^\dag$ create conduction and more localized electrons in $f$ orbit with momentum $\mathbf{k}$ and spin $\sigma$ respectively. $\xi_{\mathbf k}$ is the energy  dispersion due to the nearest hopping amplitude $t$ and is equal to $\varepsilon_{\mathbf k}-\mu$ with $\varepsilon_{\mathbf k} =-2t\sum_{i=x,y,z} \cos k_i$ and $\mu$ being the chemical potential. $\xi^d_{\mathbf{k}} = \varepsilon_d - \eta \varepsilon_{\mathbf k} -\mu$ characterizes the narrow band formed by $d$ electrons with $\eta$ being the bandwidth and $\varepsilon_{d}$ being the relative shift of band center. 
$H_{SO}$ is the spin-orbit interaction and is generally given by
\begin{eqnarray} \label{SO}
H_{SO} = \sum_{\mathbf{k} \sigma \sigma' } (2\lambda_{\mathbf{k}}^{\sigma \sigma'}  c_{\mathbf{k} \sigma}^\dag c_{\mathbf{k} \sigma'} + 2\bar{\lambda}_{\mathbf{k}}^{\sigma \sigma'} d_{\mathbf{k} \sigma}^\dag d_{\mathbf{k} \sigma'}), \label{SO}
\end{eqnarray}
where $\lambda_{\mathbf{k}}^{\sigma \sigma'} $ and $\bar{\lambda}_{\mathbf{k}}^{\sigma \sigma'}$ can be either Dirac-type spin-orbit interaction, $\boldsymbol{\sigma} \cdot \sin \mathbf{k}$\cite{Coleman}, 
or Rashba-type interaction, $\hat{z}\cdot\boldsymbol{\sigma} \times \sin \mathbf{k}_{2D}$. Here  $\sin \mathbf{k}$ denotes $(\sin k_{x},\sin k_{y},\sin k_{z})$ while
$\sin \mathbf{k}_{2D}$ denotes $(\sin k_{x},\sin k_{y},0)$. $\mathbf{V}_{\mathbf{k}}$ is the hybridization matrix (taken to be real) that describes the hybridization between $c$ and $d$ electrons and will be taken in the form $v_0+\mathbf{V}_{SO}(\mathbf{k})$ with $\mathbf{V}_{SO}(\mathbf{k})$ being   due to spin-orbit interaction and being linear in $\mathbf{k}$\cite{Coleman,Mou}. Finally, $U$ describes the Hubbard repulsion between $d$ electrons. 

In order to access the electronic structures in the large $U$ limit, the slave boson method is employed. In this method, the creation operators of $d$ electrons are represented by $d_{i\sigma}^{\dag}=f_{i\sigma}^{\dag}b_i$, where $f_{i \sigma}$ and $b_i$ are the spinon and holon operators respectively,  which satisfy the constraint $\sum_{\sigma}f_{i\sigma}^{\dag}f_{i\sigma}+b_{i}^{\dag}b_{i}=1$. This constraint can be satisfied by introducing a Lagrangian field $\lambda_{i}$. In the low temperature limit, we apply the mean-field approximation by assuming holons condense so that $\langle b_{i} \rangle=\langle b_{i}^{\dag} \rangle=r$ and $\lambda_{i}$ is replaced by its mean value $\lambda$. Consequently, the Hamiltonian becomes $H_{M} =\sum_{\mathbf{k} \sigma}  \left( c_{\mathbf{k} \sigma}, f_{\mathbf{k} \sigma} \right)^\dag h_{\mathbf{k}} \left(  c_{\mathbf{k} \sigma}, f_{\mathbf{k} \sigma} \right) + N \lambda (r^2-1)$ with
\begin{equation}\label{hk}
h_{\mathbf{k}}=
\begin{pmatrix}
\xi_{\mathbf{k}} \mathbb{1} + \lambda_{\mathbf{k}}  & \widetilde{V}_{\mathbf{k}}\\
\widetilde{V}_{\mathbf{k}}&\tilde{\xi}_{\mathbf{k}}^{d} \mathbb{1} +  r^2 \bar{\lambda}_{\mathbf{k}} \\
\end{pmatrix}.
\end{equation}
Here  $\widetilde{V}_{\mathbf{k}}=rV_{\mathbf{k}}$, $\tilde{\xi}^d_{\mathbf{k}}=(\varepsilon_d+\lambda)-\eta r^2\varepsilon_{\mathbf{k}}-\mu$, $N$ is number of lattice points, and we have made use of $\sum_{\sigma}f_{i\sigma}^{\dag}f_{i\sigma} =  \sum_{\sigma} d_{i \sigma}^\dag d_{i \sigma}$. Given the Hamiltonian $h_{\mathbf{k}}$, $\lambda$ and $r$ are determined by minimizing the free energy with respect to $\lambda$ and $r$.  As a result,  we find that $\lambda$ and $r$ can be determined by solving the following mean-field equations self-consistently
\begin{equation}
\frac{1}{N} \sum_{\mathbf{k} \sigma}  \langle f_{\mathbf{k} \sigma}^{\dag}   f_{\mathbf{k} \sigma}  \rangle
+r^2=1, \label{meanM1}
\end{equation}
\begin{equation}
\frac{1}{N} \sum_{\mathbf{k} \sigma \sigma'}   \left[ {\rm Re}  \left(V^{\sigma \sigma'}_{\mathbf {k}} \langle c_{\mathbf{k} \sigma}^{\dag}  f_{\mathbf{k} \sigma'} \rangle \right) - r (2\bar{\lambda}_{\mathbf{k}}^{\sigma \sigma'}+ \eta \varepsilon_{\mathbf{k}} \delta _{\sigma\sigma'}  ) \langle f_{\mathbf{k} \sigma}^{\dag} f_{\mathbf{k} \sigma'}\rangle  \right]+r\lambda=0. \label{meanM2}
\end{equation}
For the further analysis of the energy band, it is convenient to rewrite $\xi_{\mathbf{k}}=-\mu_{\mathbf{k}}+m_{\mathbf{k}}$, and $\tilde{\xi}_{\mathbf{k}}^{d}=-\mu_{\mathbf{k}}-m_{\mathbf{k}}$ with $\mu_{\mathbf{k}}=\mu-[(1-\eta r^{2})\varepsilon_{\mathbf{k}}+\varepsilon_{d}+\lambda]/2$, and $m_{\mathbf{k}}=[(1+\eta r^{2})\varepsilon_{\mathbf{k}}-\varepsilon_{d}-\lambda]/2$ so that the Hamiltonian can be cast in the tensor-product form as
\begin{equation}\label{H}
h_{\mathbf{k}}=-\mu_{\mathbf{k}}\tau_{0}\otimes \sigma_{0}+m_{\mathbf{k}} \tau_{z}\otimes \sigma_{0}+ \tau_{x}\otimes \widetilde{V}_{\mathbf{k}}+
(\tau_0+\tau_z) \otimes \lambda_{\mathbf{k}} +r^2(\tau_0 -\tau_z) \otimes \bar{\lambda}_{\mathbf{k}},  
\end{equation}
where $\tau_0=\sigma_0=\mathbb{1}$, $\boldsymbol{\tau}=(\tau_{x},\tau_{y},\tau_{z})$ are the Pauli matrices that act on the orbital degree ($c$ or $f$) of freedom, and $\boldsymbol{\sigma}=(\sigma_{x},\sigma_{y},\sigma_{z})$ act on the real spin. In the above orbital and spin basis, the corresponding time-reversal operator $\Theta$ and inversion operator $P$ are given by
\begin{eqnarray}
\Theta&=&i\tau_{0}\otimes\sigma_{y}K,  \\ \nonumber
P&=&\tau_{z}\otimes\sigma_{0},
\end{eqnarray}
where $K$ stands for complex conjugation\cite{Nagaosa}. It is then straightforward to see that in the absence of spin-orbit interactions ($H_{SO} =0$) and when $\mathbf{V}_{\mathbf{k}} = \mathbf{V}_{SO} (\mathbf{k})$, the Hamiltonian $h_{\mathbf{k}} $ satisfies $\Theta h_{\mathbf{k}}\Theta^{-1}=h_{\mathbf{-k}}$ and $P h _{\mathbf{k}}P^{-1}=h_{\mathbf{-k}}$. Hence the Anderson lattice model without spin-orbit interactions  is both time-reversal symmetric and inversion symmetric.
 It has been shown that this system supports stable finite-temperature Dirac points protected by both time-reversal symmetry (TRS) and inversion symmetry (IS)\cite{Mou}.
\section{Inversion symmetric Weyl semi-metallic phase}
We first consider the Weyl semi-metallic phase when the Anderson lattice is inversion symmetric. In this case, we take $H_{SO}=0$ and the hybridization matrix takes the
following form 
\begin{eqnarray}
\mathbf{V}_{\mathbf{k}} = 2\lambda_{so} \boldsymbol{\sigma} \cdot \sin \mathbf{k}.
\end{eqnarray}
The resulting Anderson model describes SmB$_6$ in which $v_0$ vanishes due to odd parity of the $f$ orbits\cite{Coleman} so that the spin-orbit interaction $\lambda_{so}$ dominates. To obtain Weyl semi-metallic phase, we further include exchange fields that break the time-reversal symmetry so that the following additional Hamiltonian is included
\begin{equation}
H_{\mathbf{k}}^{M}=\sum_{\mathbf{k}} \mathbf{M}_{c}\cdot (c_{\mathbf{k}\alpha}^{\dag} \boldsymbol{\sigma}_{\alpha \beta} c_{\mathbf{k}\beta})+\mathbf{M}_{f}\cdot (f_{\mathbf{k}\alpha}^{\dag} \boldsymbol{\sigma}_{\alpha \beta} f_{\mathbf{k}\beta}).
\end{equation}
Here $\mathbf{M}_{c}$ and $\mathbf{M}_{f}$ are exchange fields for the conduction and $d$ electrons respectively, and we have made use of
the relation $d_{\mathbf{k}\alpha}^{\dag} \boldsymbol{\sigma}_{\alpha \beta} d_{\mathbf{k}\beta} = f_{\mathbf{k}\alpha}^{\dag} \boldsymbol{\sigma}_{\alpha \beta} f_{\mathbf{k}\beta}$. For simplicity, we shall set $\mathbf{M}_{c}=\mathbf{M}_{f}=\mathbf{M}=M_{z}\hat{z}$. The resulting Hamiltonian is given by
\begin{eqnarray}\label{HM}
h_{\mathbf{k}}=-\mu_{\mathbf{k}}\tau_{0}\otimes \sigma_{0}+m_{\mathbf{k}} \tau_{z}\otimes \sigma_{0}+2r\lambda_{so} \tau_{x}\otimes  \boldsymbol{\sigma} \cdot \sin \mathbf{k} +M_{z}\tau_{0}\otimes \sigma_{z}.
\end{eqnarray}
\begin{figure}[ht]
\begin{center}
\includegraphics[height=2.3in,width=3.2in] {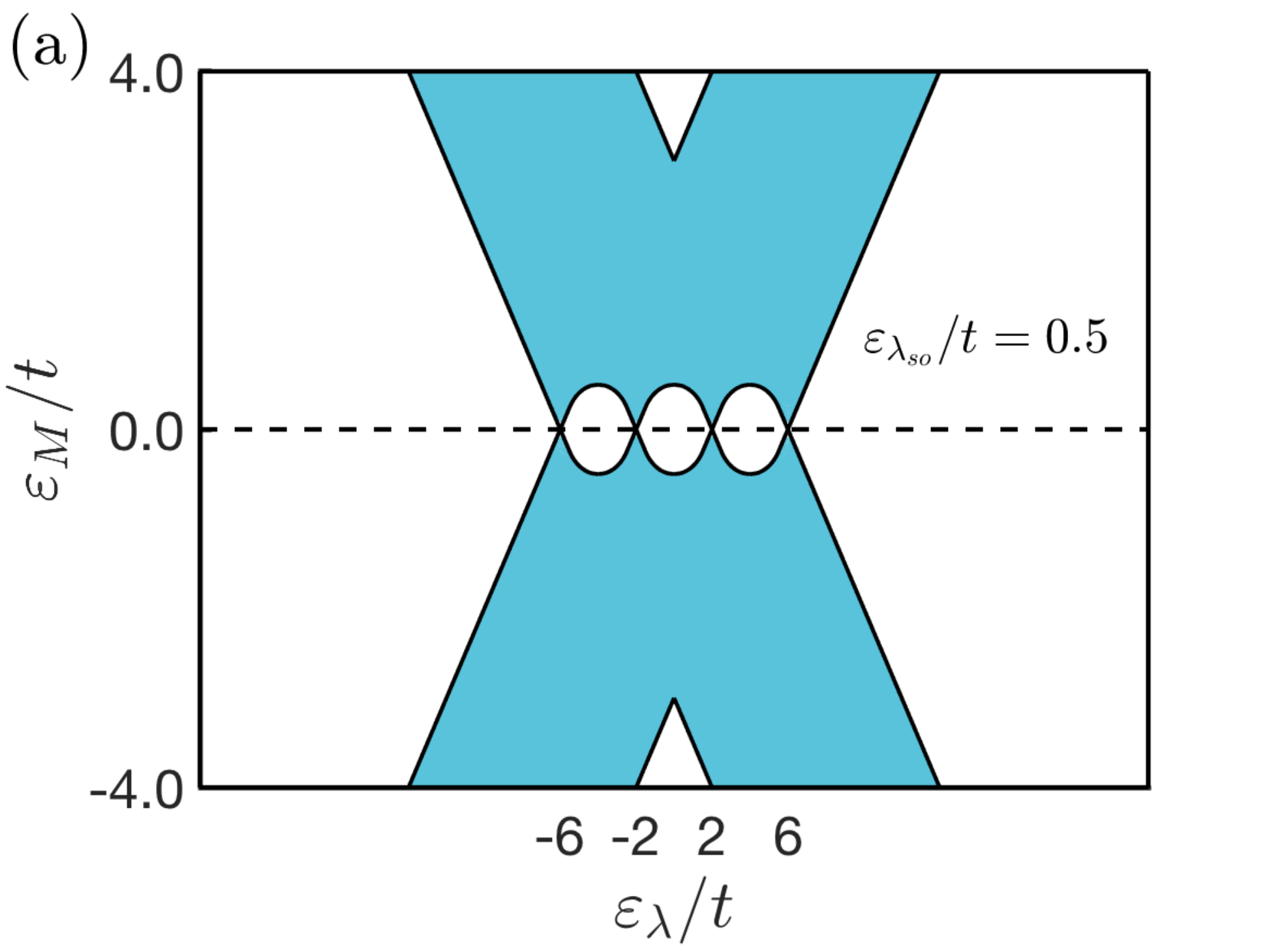}
\includegraphics[height=2.3in,width=3.2in] {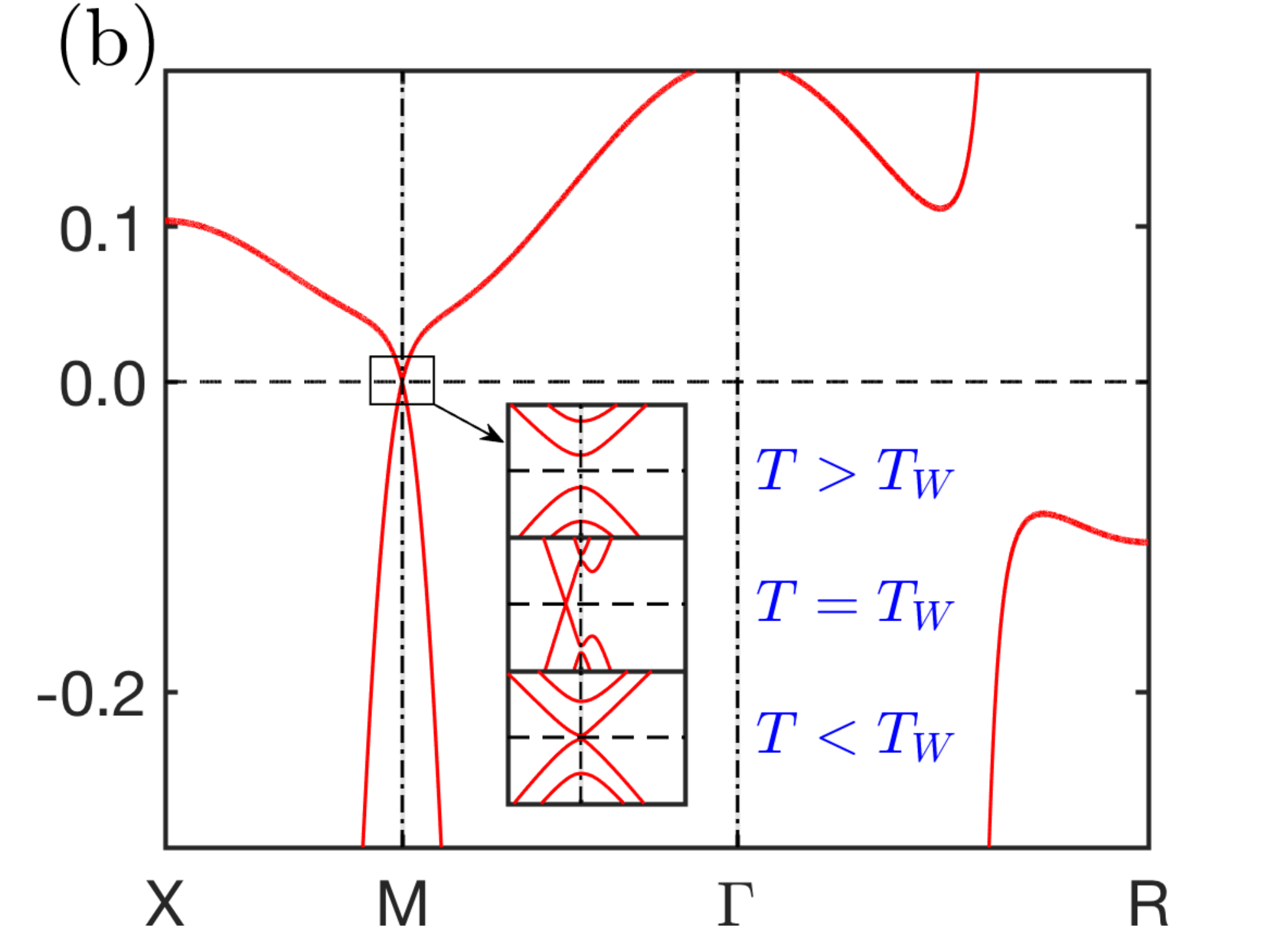}
\end{center}
\caption{(a) Phase diagram of the Anderson lattice model with an external magnetic field. Here $\varepsilon_{\lambda}  \equiv (\varepsilon_d+\lambda)/(1+\eta r^2)$, $\varepsilon_{\lambda_{so}}  \equiv r\lambda_{so}/(1+\eta r^2)$ and $\varepsilon_{M}  \equiv M/(1+\eta r^2)$. Shaded regime is the Weyl semi-metallic phase while white regimes are insulating phases with gaps in electronic structures. When $\varepsilon_{M}=0$ the gapless phases at $\varepsilon_{\lambda}/t=-6,-2,2,6$ are Dirac semi-metallic phases with corresponding Dirac points being at time reversal momenta $\Gamma, X, M, R$ respectively. (b) Emergence of a finite temperature Weyl point that splits off from $M$ point. Here $t=1$, $\lambda_{so}=0.14$, $\eta=0.05$, $\varepsilon_{d}=1.837$, $M_{z}$=0.001. The transition temperature when the Kondo insulator becomes a Weyl semi-metal occurs at $T_{W}=0.03$.}
\label{fig1}
\end{figure}
Clearly, we have $\Theta\mathcal{H}_{\mathbf{k}}\Theta^{-1}\neq\mathcal{H}_{\mathbf{-k}}$, and $P\mathcal{H}_{\mathbf{k}}P^{-1}=\mathcal{H}_{\mathbf{-k}}$. Hence the time-reversal symmetry of the system is broken while the inversion symmetry still holds. The energy spectrum  has an analytic form and is given by
\begin{equation}
E_{\mathbf{k}}^{(\alpha,\beta)}=-\mu_{\mathbf{k}} +\alpha \sqrt{(\sqrt{m_{\mathbf{k}}^2+4r^2\lambda_{so}^2\sin^2k_{z}}+\beta M_{z})^{2}+4r^2\lambda_{so}^2\sin^2\mathbf{k}_{2D}},
\end{equation}
where  $\alpha=(+,-)$ and $\beta=(+,-)$ are indices for signs. It is clear that the branch, $E_{\mathbf{k}}^{(\alpha,+)}$, is always gapful when $M_{z}>0$, while the branch, $E_{\mathbf{k}}^{(\alpha,-)}$, is always gapful when $M_{z}<0$. Obviously, gapless phases are determined by the condition $E_{\mathbf{k}}^{(+,-)}-E_{\mathbf{k}}^{(-,-)}=2\sqrt{(\sqrt{m_{\mathbf{k}}^2+4r^2\lambda_{so}^2\sin^2k_{z}}- M_{z})^{2}+4r^2\lambda_{so}^2\sin^2\mathbf{k}_{2D}}=0$. Hence by setting $\sin^2\mathbf{k}_{2D}=0$ and $\sqrt{m_{\mathbf{k_{\text{w}}}}^2+4r^2\lambda_{so}^2\sin^2k_{{\text{w}_{z}}}}- M_{z}=0$, we determine all possible gapless momenta $\mathbf{k_{\text{w}}}=(k_{\text{w}_{x}},k_{\text{w}_{y}},k_{\text{w}_{z}})$ which satisfy
\begin{equation}\label{Mphaseeq}
(\varepsilon_{\mathbf{k}_{\text{w}}}-\varepsilon_{\lambda})^{2}+16\varepsilon_{\lambda_{so}}^{2}\sin^{2}k_{\text{w}_{z}}=4\varepsilon_{M}^{2},
\end{equation}
where relevant parameters  are given by $\varepsilon_{\lambda} \equiv (\varepsilon_{d}+\lambda)/(1+\eta r^{2})$, $\varepsilon_{\lambda_{so}} \equiv r\lambda_{so}/(1+\eta r^{2})$, and $\varepsilon_{M} \equiv M_{z}/(1+\eta r^{2})$. It is clear to see that $\varepsilon_{\lambda}$, $\varepsilon_{\lambda_{so}}$ and $\varepsilon_{M}$ are the effective parameters that tune the Kondo insulator into different phases. Solutions to Eq.(\ref{Mphaseeq}) give rise to phase diagrams shown in Fig. \ref{fig1}(a), where the gapless Weyl semi-metallic phase is shown as the shaded regime.  Furthermore, by solving mean-field equations, Eqs.(\ref{meanM1}) and (\ref{meanM2}), we find that it is possible to tune the Kondo insulator so that it becomes a Weyl semi-metal at finite temperatures.  As shown in FIg. \ref{fig1}(b), the transition occurs at $T_W=0.03$ when parameters are taken at $t=1$, $\lambda_{so}=0.14$, $\eta=0.05$, $\varepsilon_{d}=1.837$, $M_{z}=0.001$.

\begin{figure}[ht]
\includegraphics[height=2.3in,width=3.2in] {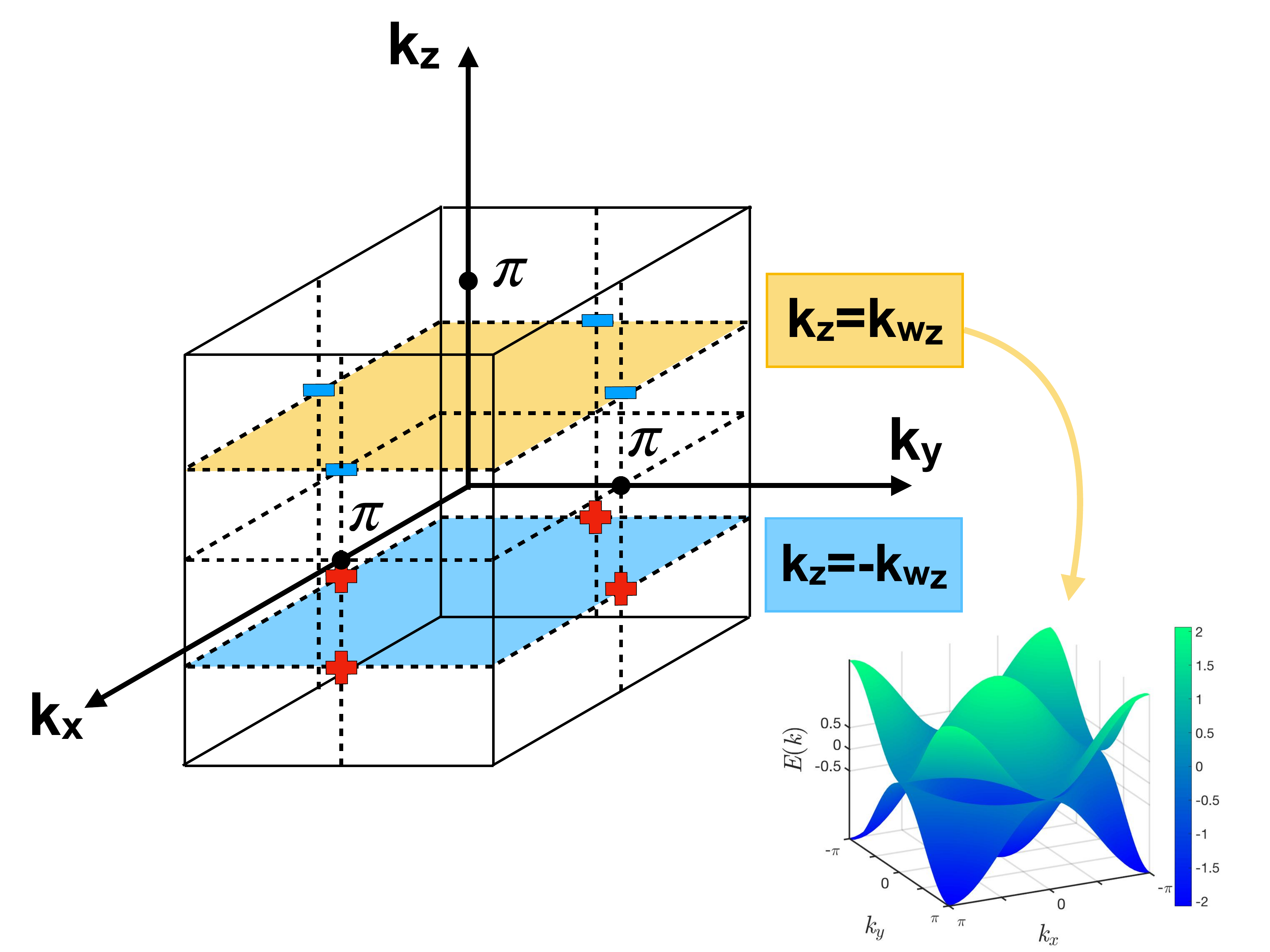}
\includegraphics[height=2.3in,width=3.2in] {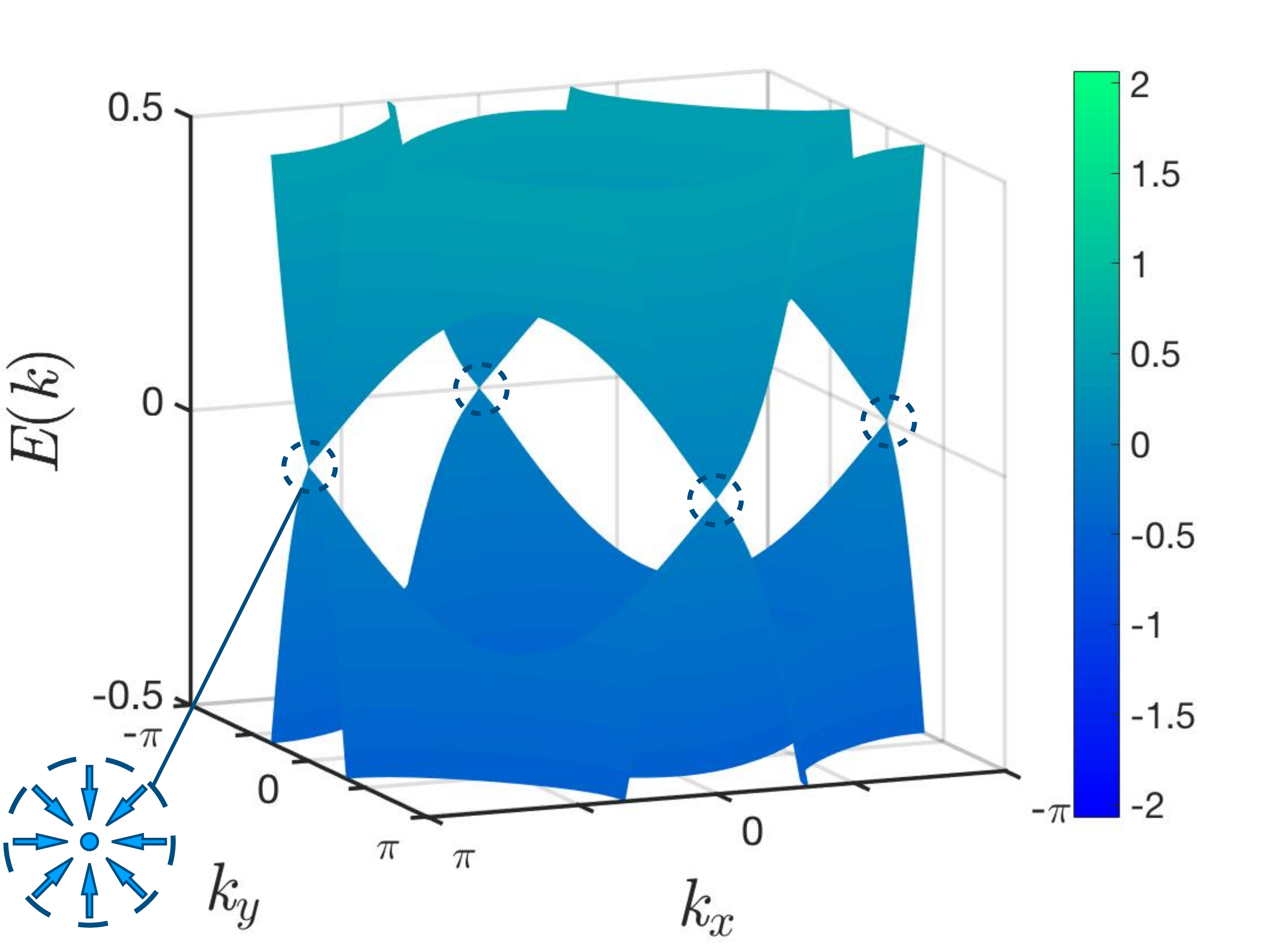}
\caption{Sketch of the distribution of Weyl nodes in the first Brillouin zone. The monopole charges of each Weyl nodes are denoted by  $-$ or $+$. Weyl nodes move along dash lines when temperature changes. Notice that in order to display Weyl nodes clearly, here $E(k)$ does not include the effective chemical potential $\mu_{\mathbf{k}}$.}
\label{fig2}
\end{figure}

Here we further analyze charge associated with the Weyl point located at $\mathbf{k_{\text{w}}}=(0,\pi,k_{\text{w}_{z}})$. Near the nodal point, the linearized Hamiltonian can be re-casted into the form 
\begin{equation}\label{chargematrix}
h_{\mathbf{k_{\text{w}}}+\mathbf{q}}=
\begin{pmatrix}
\xi_{\mathbf{k_{\text{w}}}+\mathbf{q}}^{+}&0*\mathbbm{1}_{2\times2}\\
0*\mathbbm{1}_{2\times2}&\xi_{\mathbf{k_{\text{w}}}+\mathbf{q}}^{-}\\
\end{pmatrix}
\end{equation}
with
\begin{equation}\label{HBcharge}
\xi_{\mathbf{k_{\text{w}}}+\mathbf{q}}^{\pm}\equiv -\mu_{\mathbf{k_{\text{w}}}+\mathbf{q}}\tau_{0}+(\widetilde{m}_{\mathbf{k_{\text{w}}}+\mathbf{q}}\pm M_{z})\tau_{z}+2r\lambda_{so}q_{\pm}\tau_{+}+\text{h.c.},
\end{equation}
where $\tau_{\pm}=(\tau_{x} \pm i\tau_{y})/2$, $q_{\pm}=q_{x}\pm iq_{y}$, and $\widetilde{m}_{\mathbf{k_{\text{w}}}+\mathbf{q}}=\sqrt{m_{\mathbf{k_{\text{w}}}+\mathbf{q}}^{2}+4r^{2}\lambda_{so}^{2}(q_{z}\cos k_{\text{w}_{z}})}$.  It is clear that $\xi_{\mathbf{k_{\text{w}}}+\mathbf{q}}^{+}$ is always gapful, while $\xi_{\mathbf{k_{\text{w}}}+\mathbf{q}}^{-}$ can be tuned into gapless regime. From the linearized Hamiltonian, we identify the net monopole charge associated the Weyl node at  $(0,\pi,k_{\text{w}_{z}})$ is $-1$\cite{Bernevig}. Similar analysis allows one to identify all charges of Weyl nodes. This is sketched in Fig.~\ref{fig2}.

\section{Weyl nodal-ring semi-metallic phase}
\begin{figure}[ht]
\includegraphics[height=2.5in,width=3.1in] {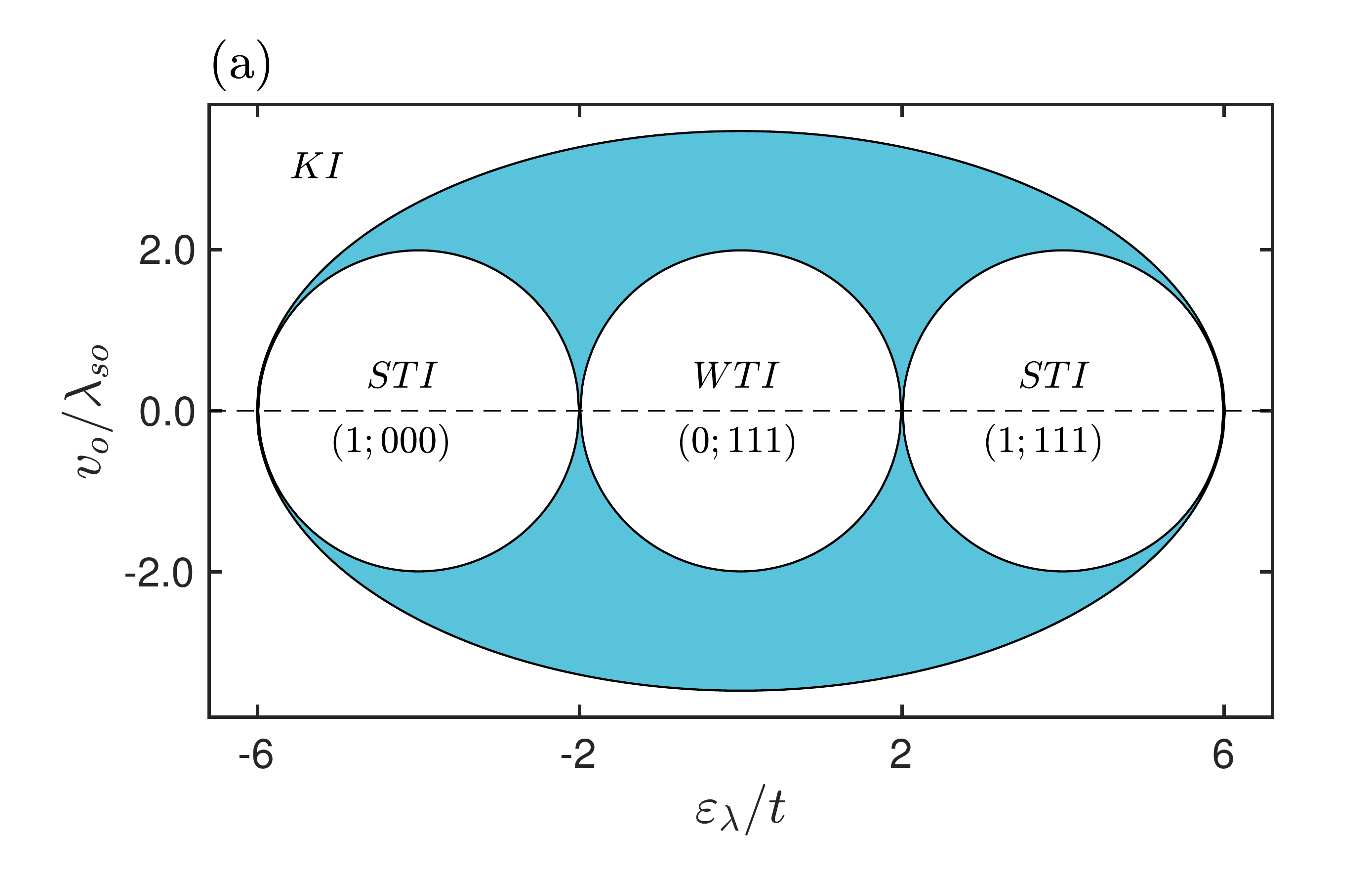}
\includegraphics[height=2.5in,width=3.1in] {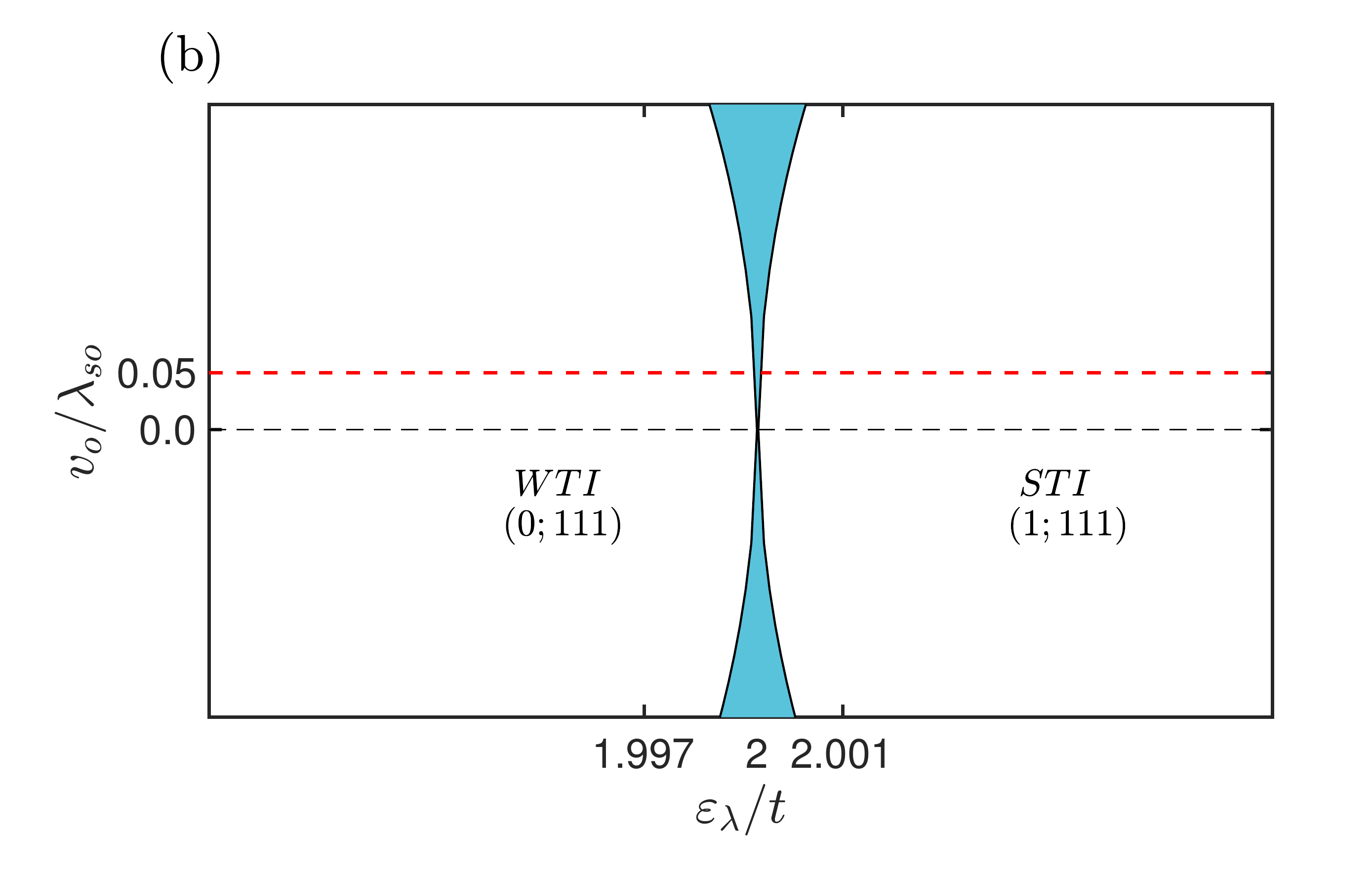}
\caption{(a) Topological phase diagram in the presence of constant hybridization $v_{0}$ between $c$ and $f$ electrons, shaded regimes indicate gapless phases while white regimes being insulating phases, labelled by strong topological insulator (STI), weak topological insulator (WTI), and Kondo insulator (KI) when the valence bands are filled\cite{Mou}. Here $\varepsilon_{\lambda}  \equiv \frac{\varepsilon_d+\lambda}{1+\eta r^2}$ and $(\nu_0;\nu_1, \nu_2,\nu_3)$ are topological indices. (b) The phase transition path (the red dash line) in the phase diagram with $t=1$, $\varepsilon_{d}=1.724$, $v_{0}=0.01$, $\lambda_{so}=0.2$. When the temperature increases from $T=0.02$ to $T=0.04$, the Anderson lattice goes through Weyl nodal-ring semi-metallic phases.}
\label{fig3}
\end{figure}

In this section, we demonstrate that the Weyl nodal-ring semi-metallic phases emerge in the Anderson lattice model when inversion symmetry is broken. 
In this case, we take $H_{SO}=0$ and the hybridization matrix takes the following form\cite{Mou}
\begin{eqnarray}
\mathbf{V}_{\mathbf{k}} = v_0+2\lambda_{so} \boldsymbol{\sigma} \cdot \sin \mathbf{k}.
\end{eqnarray}
After applying the slave-boson approximation, the mean-field Hamiltonian is given by
\begin{equation}\label{HV0}
h_{\mathbf{k}}=-\mu_{\mathbf{k}}\tau_{0}\otimes \sigma_{0}+m_{\mathbf{k}} \tau_{z}\otimes \sigma_{0} +r \tau_{x}\otimes ( v_{0}\sigma_{0}+2\lambda_{so}\boldsymbol{\sigma} \cdot \sin \mathbf{k}).
\end{equation}
The energy spectrum to $h_{\mathbf{k}}$ has an analytic form and is given by
\begin{equation}
E_{\mathbf{k}}^{(\alpha,\beta)}=-\mu_{\mathbf{k}}+\alpha\sqrt{m_{\mathbf{k}}^{2}+r^{2}\left(v_{0}+2\beta\lambda_{so}\sqrt{\sin^{2}\mathbf{k}}\right)^{2}},
\end{equation}
where $\alpha$ and $\beta$ are $+$ or $-$. Here it is clear that the gap is determined by $E_{\mathbf{k}}^{(+,\beta)}-E_{\mathbf{k}}^{(-,\beta)}=2\sqrt{m_{\mathbf{k}}^{2}+r^{2}\left(v_{0}+2\beta\lambda_{so}\sqrt{\sin^{2}\mathbf{k}_{3D}}\right)^{2}}$. Obviously, when $\beta=+$, $E_{\mathbf{k}}^{(\alpha,\beta)}$ remains gapful. Furthermore, gapless points,  $\mathbf{k}_{0}$,  are determined by setting $m_{\mathbf{k}_{0}}=0$, and $v_{0}-2\lambda_{so}\sqrt{\sin^{2}\mathbf{k}_{0}}=0$. These two equations are equivalent to
\begin{equation}\label{ps}
\sum_{i=x,y,z}\cos k_{i}=-\frac{\varepsilon_{\lambda}}{2t},\ \ \sum_{i=x,y,z}\cos^{2} k_{i}=3-\left(\frac{v_{0}}{2\lambda_{so}}\right)^{2}.
\end{equation}
The solution, $u_i=\cos k_{i}$, to the second equation in Eq.(\ref{ps}) forms a sphere with radius, $\sqrt{u_x^2+u^2_y+u^2_z}$, equals to $\sqrt{3-\left(v_{0}/2\lambda_{so}\right)^{2}}$, while the first equation represents a plane. The distance between the center of the sphere and the plane is given by $\left| \varepsilon_{\lambda}/2\sqrt{3}t\right|$ so that Eq.(\ref{ps}) has solutions only if $\left| \varepsilon_{\lambda}/2\sqrt{3}t\right| \leq \sqrt{3-\left(v_{0}/2\lambda_{so}\right)^{2}}$. The solutions of Eq.(\ref{ps}) form curves as illustrated as the boundaries of shaded area in Fig \ref{fig3}(a). 
Right at the boundary, $\left| \varepsilon_{\lambda}/2\sqrt{3}t\right|  = \sqrt{3-\left(v_{0}/2\lambda_{so}\right)^{2}}$, the plane and sphere touches at a point,
which gives rises to Weyl nodes. The system is thus a Weyl semi-metal. However, when $\left| \varepsilon_{\lambda}/2\sqrt{3}t\right| < \sqrt{3-\left(v_{0}/2\lambda_{so}\right)^{2}}$ the intersection of a plane and a sphere is a ring in $k$ space. Hence, instead of being Weyl semi-metallic phases, 
we find that Weyl nodal-ring semi-metallic phases emerge inside the shaded regime in Fig \ref{fig3}(a).

The Weyl nodal-ring lies in the surface defined by $m_{\mathbf{k}}=0$. Following Ref.\cite{Nagaosa}, near the center of the ring on the surface, by performing the expansion of the wave-vector in the local frame to linear terms and removing the smooth energy background term, we obtain that the effective Hamiltonian is given by 
\begin{equation}\label{eff}
h_{eff} (\mathbf{k}' )= r  v_{0} \tau_{x} \otimes \sigma_{0}+2r \lambda_{so} \tau_{x} \otimes (\sigma_x {k_x'}+ \sigma_z  k_z'),   
\end{equation}
where $\mathbf{k}'=(k_x',k_y',k_z')$ with $k_x'$and $k_z'$ being the components in parallel and in perpendicular to the surface defined by $m_{\mathbf{k}}=0$ respectively. Here the local coordinates are chosen such that the  components in parallel to the surface defined by $m_{\mathbf{k}}=0$ is aligned to the $k_x'$ axis. The Hamiltonian $h_{eff}$ is mirror symmetric 
\begin{equation}
M^{-1} h_{eff} (k_x',k_y',-k_z') M = h_{eff} (k_x',k_y',k_z'),   
\end{equation}
where $M= \tau_x \otimes i \sigma_x$ is the corresponding representation of the mirror symmetry operator. The system is thus mirror symmetric with respect to the surface defined by $m_{\mathbf{k}}=0$. Since the inversion symmetry is broken when $v_0$ is non-vanishing, the Weyl nodal-ring emerges as the consequence of the presence of mirror symmetry and the broken inversion symmetry\cite{Yang}.
In addition, the effective Hamiltonian is time-reversal invariant and has particle-hole symmetry with the charge conjugation being given by $C = -\tau_y \otimes i \sigma_y K$. The nodal-ring is protected by these symmetries and belong to the class CII with $R_{+-}$ defined in Ref.\cite{Chiu}.

The Anderson lattice is tunable in temperature. In Fig.\ref{fig3}(b), we demonstrate that finite temperature phase transitions between strong topological insulating phase (STI) and weak topological insulating phase (WTI) through Weyl nodal-ring semi-metallic phase can be achieved by changing temperature. By solving the corresponding self-consistent equations, Eqs.(\ref{meanM1}) and (\ref{meanM2}), we find that there is a phase transition from STI ($\varepsilon_{\lambda}/t=2.001, T=0.02$) to WTI  ($\varepsilon_{\lambda}/t=1.997, T=0.04$) through the Weyl nodal-ring semi-metallic phase ($\varepsilon_{\lambda}/t=2, T=0.03$), as illustrated as  the red dash line shown in Fig.\ref{fig3}(b).

\section{Time reversal symmetric Weyl semi-metallic phases}
In this section, we explore the possibility of forming  the Weyl semi-metallic phases in Anderson lattice that are time-reversal symmetric. 
For this purpose, the Anderson lattice model must break the inversion symmetry. In addition, the emergence of Weyl semi-metallic phases 
requires certain crystal symmetries, in which 3D rotational symmetry breaks down to axial symmetries\cite{Bernevig}. Therefore, to realize
the Weyl semi-metallic phase with time reversal symmetry, the Anderson lattice is assumed to have layered structure with axial symmetry. 
The energy dispersion $\varepsilon_{\mathbf{k}}$ is given by
\begin{equation}
\varepsilon_{\mathbf{k}}=-2t\sum_{i=x,y,z}a_{i}\cos k_{i}
\end{equation}
where $a_{i}$ represents the relative hopping strength of each direction and we shall set $0<a_{z}<a_{y}<a_{x}=1$. In addition,
the hybridization matrix takes the two dimensional form with either 
\begin{eqnarray}
\mathbf{V}^1_{\mathbf{k}} = 2\lambda_{so} \boldsymbol{\sigma} \cdot \sin \mathbf{k}_{2D},
\end{eqnarray}
or the Rashba spin-orbit interaction
\begin{equation}
\mathbf{V}^2_{\mathbf{k}}=2 \lambda_{Ra} \hat{z}\cdot\boldsymbol{\sigma} \times \sin \mathbf{k}_{2D},
\end{equation}
where $\lambda_{so}$ and $\lambda_{Ra}$ describes the strength of different form of 2D spin-orbit interactions respectively.

We shall first consider $\mathbf{V}^1_{\mathbf{k}}$. For bulk spin-orbit interactions, in the simplest situation, 
both $c$ and $d$ electrons are governed by the same bulk spin-orbit interactions, which are characterized by setting
$\lambda_{\mathbf{k}} = \boldsymbol{\sigma} \cdot \sin \mathbf{k} $ and  $r^2 \bar{\lambda}_{\mathbf{k}} = \boldsymbol{\sigma} \cdot \sin \mathbf{k} $ 
in Eq.(\ref{H}). As a result, the total Hamiltonian is given by
\begin{eqnarray}\label{inversionH}
h_{\mathbf{k}}&=&-\mu_{\mathbf{k}}\tau_{0}\otimes \sigma_{0}+m_{\mathbf{k}} \tau_{z}\otimes \sigma_{0} +2r\lambda_{so} \tau_{x}\otimes  \boldsymbol{\sigma} \cdot \sin \mathbf{k}_{2D}  \nonumber \\
&+&\tilde{\lambda}_{so}(\tau_{0}+\tau_{z})\otimes \boldsymbol{\sigma} \cdot \sin \mathbf{k}+\tilde{\lambda}_{so}(\tau_{z}-\tau_{0})\otimes \boldsymbol{\sigma} \cdot \sin \mathbf{k}, 
\end{eqnarray}
where $\tilde{\lambda}_{so}$ describes the strength of the bulk spin-orbit interaction. In this case, the energy spectrum to Eq.(\ref{inversionH}) has an 
analytic form, which is given by
\begin{eqnarray}
& & E_{\mathbf{k}}^{(\alpha,\beta)} =-\mu_{\mathbf{k}} \\ \nonumber
&+&\beta\sqrt{m_{\mathbf{k}}^{2}+4r^{2}\lambda_{so}^{2}\sin^{2}\mathbf{k}_{2D}+4\tilde{\lambda}_{so}^{2}\sin^{2}\mathbf{k}+4\alpha\tilde{\lambda}_{so}\sqrt{m_{\mathbf{k}}^{2}\sin^{2}\mathbf{k}+4r^{2}\lambda_{so}^{2}\left|\sin\mathbf{k} \times\sin\mathbf{k}_{2D} \right|^{2}}},
\end{eqnarray}
\begin{figure}[t]
\includegraphics[height=2.4in,width=4.3in] {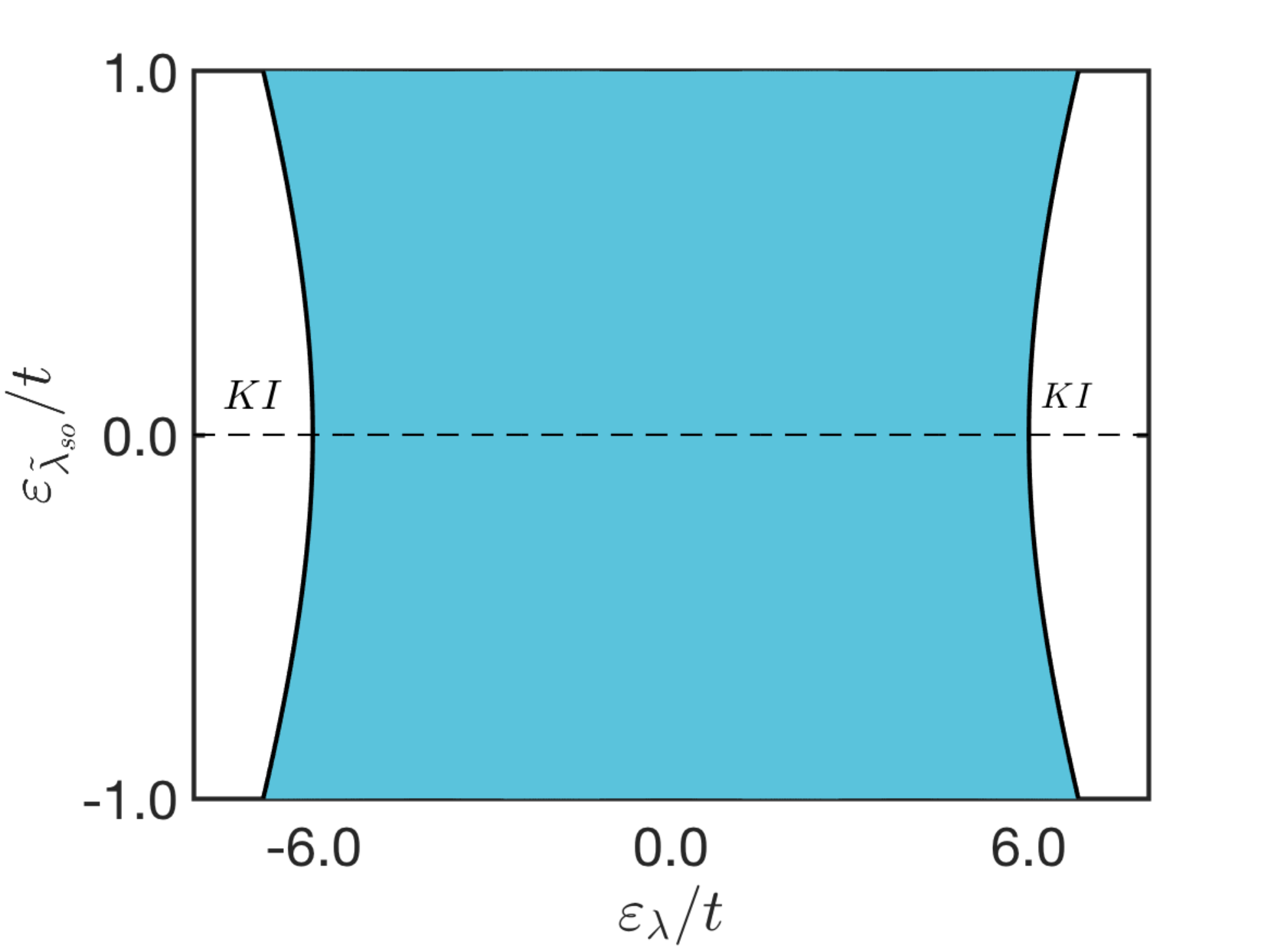}
\caption{Topological phase diagram in the presence of the bulk spin-orbit interaction in $c$ and $d$  electrons with equal strength. Here the hopping amplitude is isotropic with $a_{z}=a_{y}=a_{x}=1$. The shaded regime is the Weyl semi-metallic phases while the white regime is the Kondo insulating phase.}
\label{fig4}
\end{figure}
where $\alpha$ and $\beta$ equals to $\pm 1$. It is clear that
$E_{\mathbf{k}}^{(+,\beta)}$ remains gapfull all the times. Hence the gapless phase occurs in the branch $\alpha=-1$.  The gapless point occurs by 
requiring $\left|\sin\mathbf{k} \times\sin\mathbf{k}_{2D} \right|$ reaching its maximum value, i.e.,  $\sin \mathbf{k} \cdot \sin\mathbf{k}_{2D}=0$ and at the same time, $\sin^{2}\mathbf{k}_{2D}=0$. As a result, we find $ m_{\mathbf{k}}-2\tilde{\lambda}_{so}\sqrt{\sin^{2}\mathbf{k}}=0$ has to be satisfied. The condition for the occurrence of the Weyl semi-metallic phase is then given by
\begin{equation}\label{HSOnodph}
(\varepsilon_{\mathbf{k}_{\text{w}}}-\varepsilon_{\lambda})^{2}=4\varepsilon_{\tilde{\lambda}_{so}}^{2}\sin^{2}k_{\text{w}_{z}}
\end{equation}
where $\varepsilon_{\tilde{\lambda}_{so}}\equiv \tilde{\lambda}_{so}/(1+\eta r^{2})$. In the isotropic limit when $a_{z}=a_{y}=a_{x}=1$, solutions to Eq.(\ref{HSOnodph}) form the boundary curve between the Weyl semi-metallic phase and the Kondo insulating phase as  illustrated in Fig.~\ref{fig4}.

The monopole charge of the Weyl node in this case can be analyzed by linearizing the Hamiltonian near the node. Take the Weyl node at the point $X$, $\mathbf{k}_{\text{w}}=(0,\pi,k_{\text{w}_{z}})$, as an example, after linearized Hamiltonian can be rewritten in the form of  Eq.(\ref{chargematrix}) with
\begin{eqnarray}\label{HSOcharge}
\xi_{\mathbf{k_{\text{w}}}+\mathbf{q}}^{\pm} & = & (-\mu_{\mathbf{k_{\text{w}}}+\mathbf{q}}\pm K_{z})\tau_{0}+(m_{\mathbf{k_{\text{w}}}+\mathbf{q}}\pm K_{z})\tau_{z}+\frac{4r\lambda_{so}}{\sin k_{\text{w}_{z}}}q_{\pm}^{2}\tau_{+}+\text{h.c.} 
\end{eqnarray}
where $K_{z}=2\tilde{\lambda}_{so}(q_{z}\cos k_{\text{w}_{z}}+\sin k_{\text{w}_{z}})$. Clearly, $\xi_{\mathbf{k_{\text{w}}}+\mathbf{q}}^{+}$ is gapful while $\xi_{\mathbf{k_{\text{w}}}+\mathbf{q}}^{-}$ can be tuned into gapless phases. The monopole charge corresponding to Eq.(\ref{HSOcharge}) is $-2$\cite{Bernevig}. Hence the monopole charges of Weyl nodes in the Weyl semi-metallic phase shown in Fig.~\ref{fig4} are $\pm 2$.
\begin{figure}[t]
\includegraphics[height=2.4in,width=3.0in] {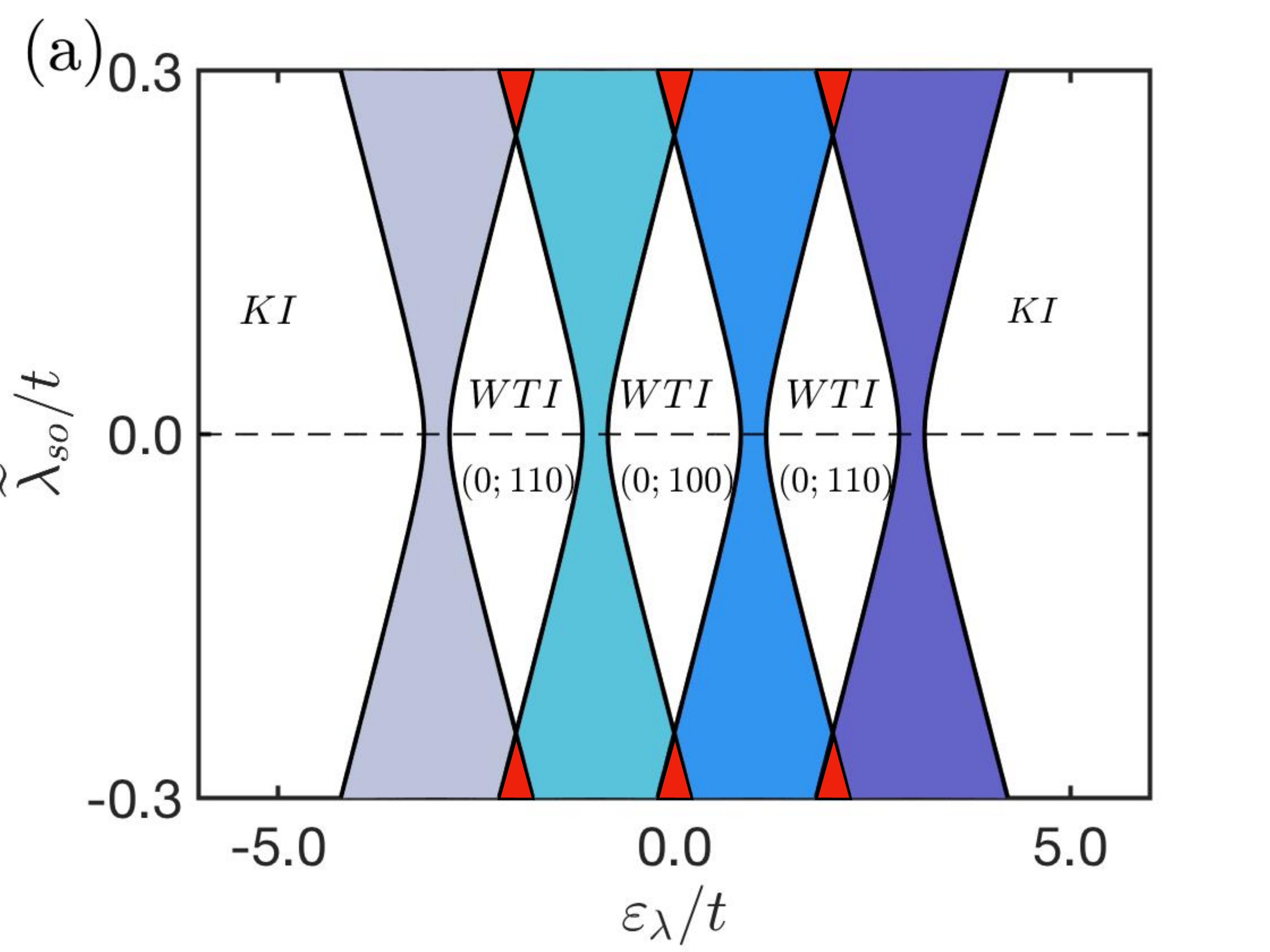}
\includegraphics[height=2.4in,width=3.0in] {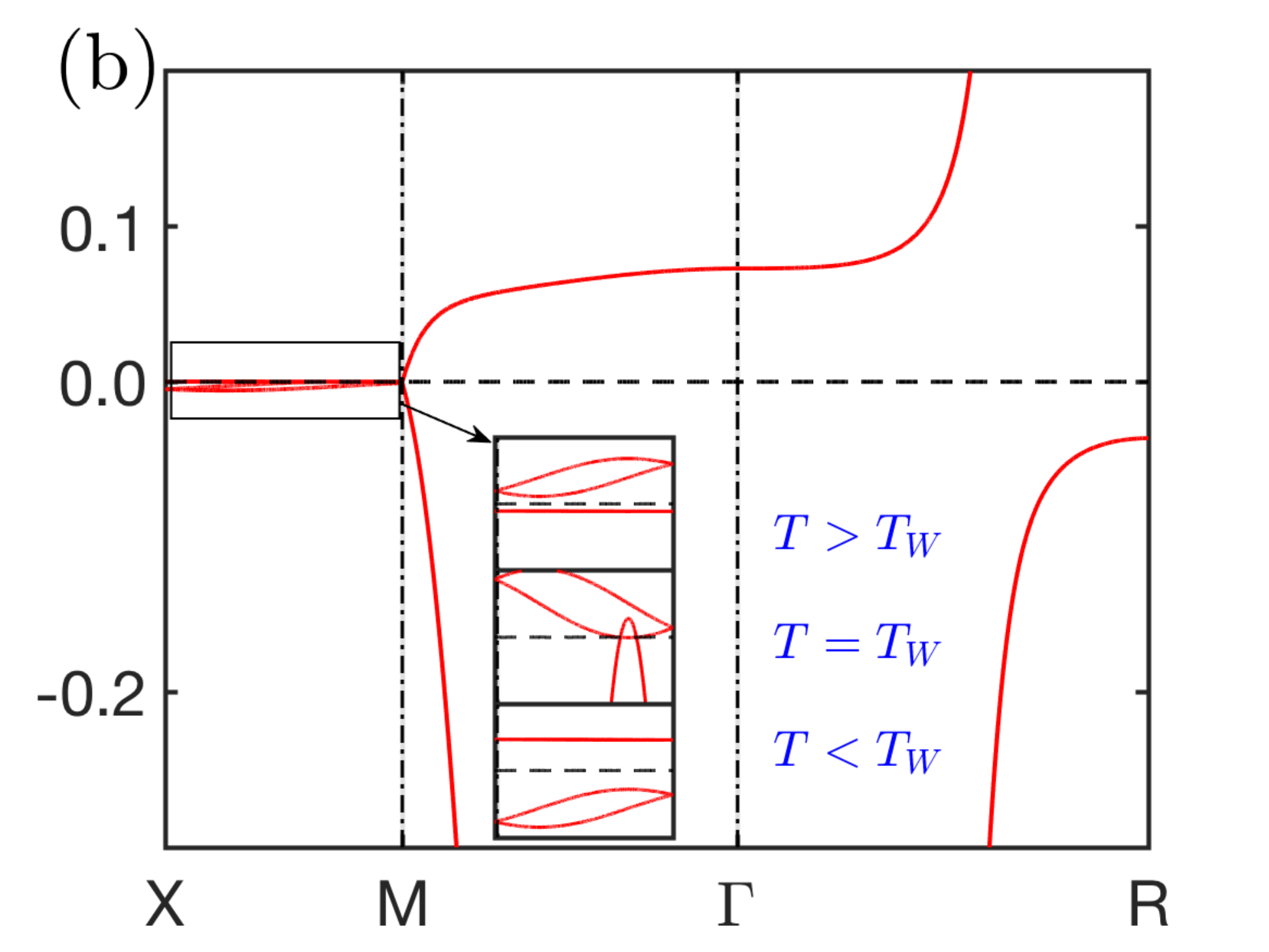}
\caption{(a)Topological phase diagram when the inversion symmetry of the Anderson lattice is broken. Here $z$ axis is the high symmetry axis and the anisotropy of hopping amplitude is characterized with $a_{x}=1$, $a_{y}=0.5$, $a_{z}=0.08$.  Here gray, green, blue, and purple regimes are Weyl semi-metallic phases with Weyl nodes emerge at $\mathbf{k}_{\text{w}}=(0,0,k_{\text{w}_{z}})$, $(0,\pi,k_{\text{w}_{z}})$,$(\pi,0,k_{\text{w}_{z}})$, and $(\pi,\pi,k_{\text{w}_{z}})$ respectively, while the red regime is the overlap regime with the emergence of both Weyl nodes from the overlapping Weyl semi-metallic phases. (b)Emergence of the  Weyl semi-metallic phase at finite temperature. Here Weyl nodes emerge in $k_z$ axis from $X$ $(\pi,0,0)$ to $M$ $(\pi,0,\pi)$.  Note that two intersecting points in the inset may look like a nodal-ring. Clear demonstration of these intersecting points being Weyl points is shown in Fig.~\ref{fig6}.
Parameters taken are  $t=1$, $\lambda_{so}=0.2$, $\eta=0.05$, $\varepsilon_{d}=0.705$, $\tilde{\lambda}_{so}=0.001$, $a_{y}=0.5$, and $a_{z}=0.001$. The critical temperature for emergence of Weyl nodes is $T_{W}=0.03$.  }
\label{fig5}
\end{figure}

\begin{figure}[h]
\includegraphics[height=2.6in,width=3.2in] {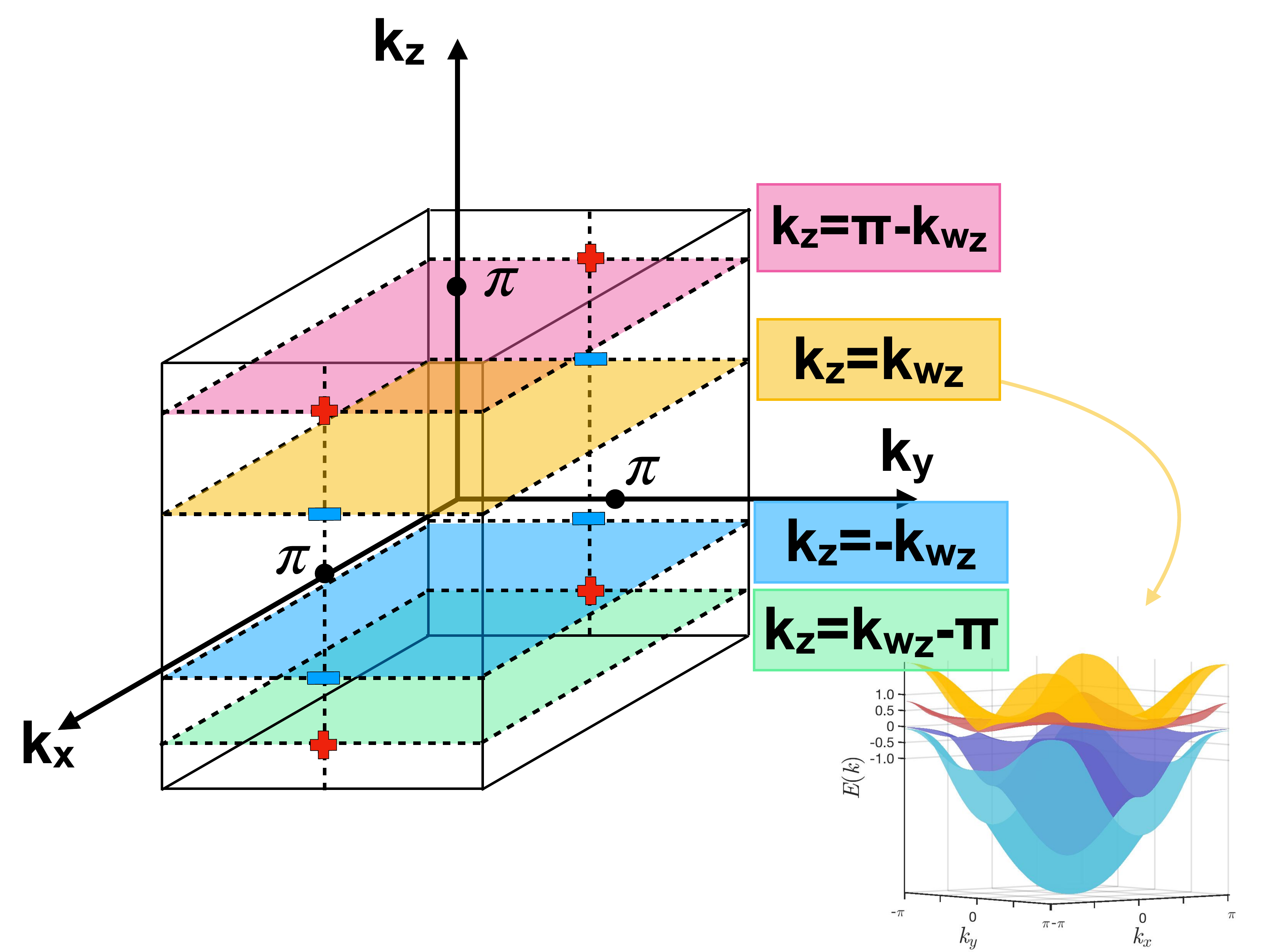}
\includegraphics[height=2.6in,width=3.2in] {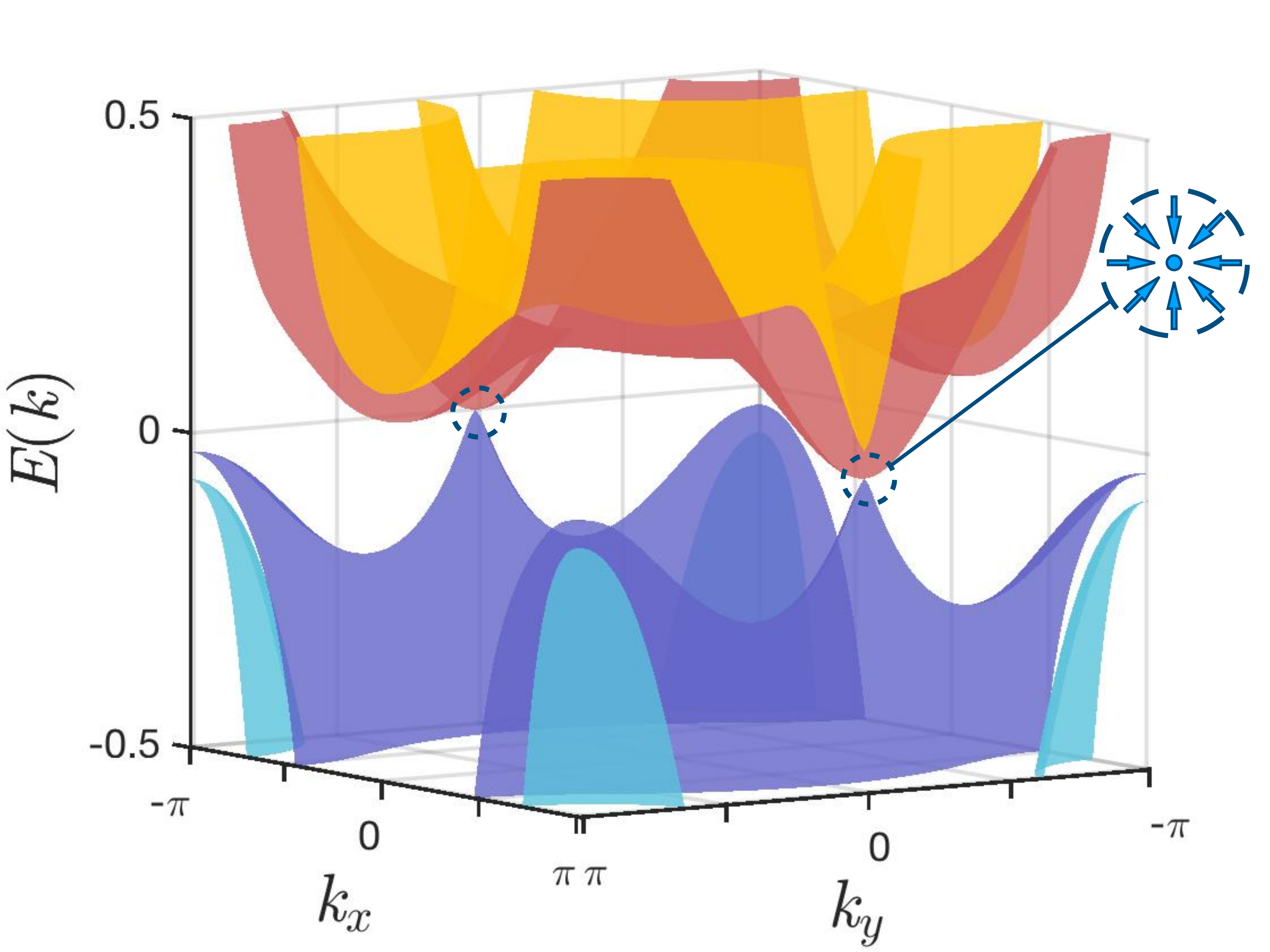}
\caption{Sketch of the distribution of Weyl nodes in the first Brillouin zone for Kondo-Weyl semimetals without inversion symmetry. The monopole charge of each Weyl nodes is $+2$ or $-2$ and  is denoted by $+$ or $-$ respectively. Weyl nodes will move along dash lines with changing parameters of the system.  Here parameters taken are $t=1$, $\lambda_{so}=0.3$, $\eta=0.05$, $\varepsilon_{d}=1.305$, $\tilde{\lambda}_{so}=0.3$, $a_{y}=0.5$, $a_{z}=0.1$, and $T= T_{W}=0.03$. Notice that in order to display Weyl nodes clearly, here $E(k)$ does not include the effective chemical potential $\mu_{\mathbf{k}}$.}
\label{fig6}
\end{figure}

In real materials, bulk spin-orbit interactions in $c$ and $d$ electrons are generally not of the same strength. Therefore, we consider a relative strength in the spin-orbit interaction of $d$ electrons. After applying the mean-field slave boson approximation, the Hamiltonian is given by
\begin{eqnarray}
h_{\mathbf{k}}&=&-\mu_{\mathbf{k}}\tau_{0}\otimes \sigma_{0}+m_{\mathbf{k}} \tau_{z}\otimes \sigma_{0} +2r\lambda_{so} \tau_{x}\otimes  \boldsymbol{\sigma} \cdot \sin \mathbf{k}_{2D}  \nonumber \\
&+&\tilde{\lambda}_{so}(\tau_{0}+\tau_{z})\otimes \boldsymbol{\sigma} \cdot \sin \mathbf{k}_{3D}+\eta r^{2}\tilde{\lambda}_{so}(\tau_{z}-\tau_{0})\otimes \boldsymbol{\sigma} \cdot \sin \mathbf{k}_{3D}. \label{spinorbit}
\end{eqnarray}
Unfortunately, the energy spectrum to $h_{\mathbf{k}}$ in Eq.(\ref{spinorbit}) no longer has an analytic form. However, because the system is axial symmetric with respect to $z$ axis and the gapless phase occurs when $\sin^{2} \mathbf{k}_{2D}=0$, the relevant spectrum for Weyl semi-metallic phase is determined by the spectrum along $z$ axis. As we can see, along $z$ axis, $E^{(\alpha,+)}(0,0,k_z)$ remains gapped. The possible gapless phases are thus determined by $E_{\mathbf{k}}^{(-,+)}-E_{\mathbf{k}}^{(-,-)} |_{\mathbf{k}=(0,0,k_z)}=2\left[m_{k_{z}}-\tilde{\lambda}_{so}(1+\eta r^{2})\sin k_{z}\right]$. Hence $m_{k_{z}}=2\tilde{\lambda}_{so}(1+\eta r^{2})\sin k_{z}$ and $\sin^{2} \mathbf{k}_{2D}=0$ determine all possible gapless phases with the corresponding nodal point $\mathbf{k}_{\text{w}} = (0,0, k_{\text{w}_{z}})$. The condition for the Weyl semi-metallic phase is then given by
\begin{equation}\label{HSOPhaseeq}
(\varepsilon_{\mathbf{k}_{\text{w}}}-\varepsilon_{\lambda})^{2}=4\tilde{\lambda}_{so}^{2}\sin^{2}k_{\text{w}_{z}}
\end{equation}
By including the anisotropy of hopping amplitudes with $\varepsilon_{\mathbf{k}}=-2t\sum_{i=x,y,z}a_{i}\cos k_{i}$ and $0<a_{z}<a_{y}<a_{x}=1$, 
solutions to Eq.(\ref{HSOPhaseeq}) form the boundary curve between the Weyl semi-metallic phase and insulating phases as  illustrated in Fig.~\ref{fig5}(a).
Here insulating phases are weak (WTI) or strong topological insulating phases (STI) labelled by the corresponding topological indices\cite{Fu-3}. Furthermore, by solving mean-field equations, Eqs.(\ref{meanM1}) and (\ref{meanM2}), we find that it is possible to tune the Kondo insulator across Weyl semi-metallic phases at finite temperatures.  As illustrated in Fig. \ref{fig5}(b), the transition occurs at $T_W=0.03$ when parameters are taken at $t=1$, $\lambda_{so}=0.2$, $\eta=0.05$, $\varepsilon_{d}=0.705$, $\tilde{\lambda}_{so}=0.001$, $a_{y}=0.5$, and $a_{z}=0.001$. It is seen that the critical temperature for emergence of Weyl nodes is $T_{W}=0.03$. In addition, as indicated by the linearized Hamiltonian in Eq.($\ref{HSOcharge}$), the net monopole charge associated each Weyl node is $\pm 2$. The distribution of Weyl nodes is sketched in Fig.~\ref{fig6}.

\begin{figure}[ht]
\includegraphics[height=2.4in,width=3.2in] {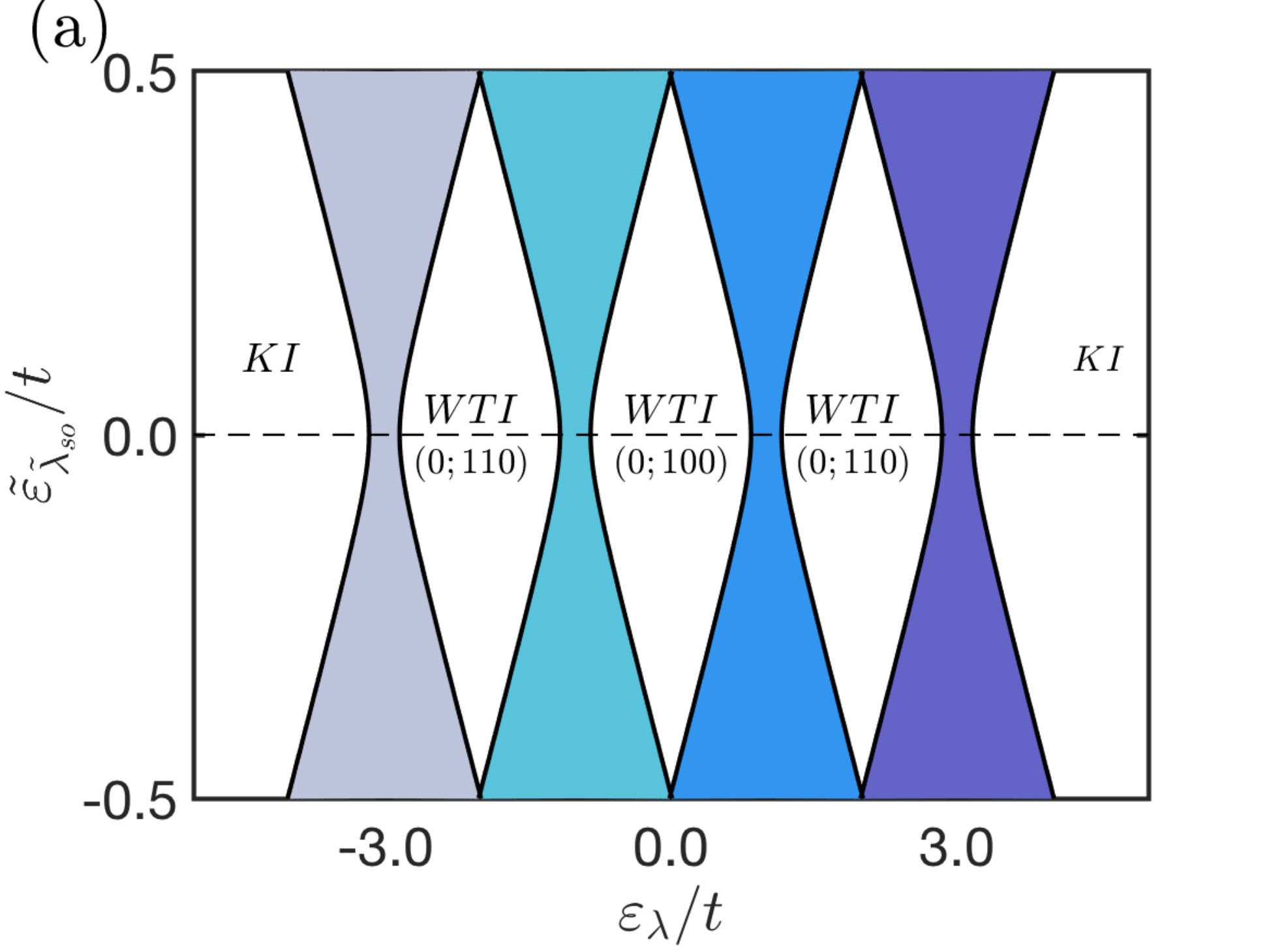}
\includegraphics[height=2.4in,width=3.2in] {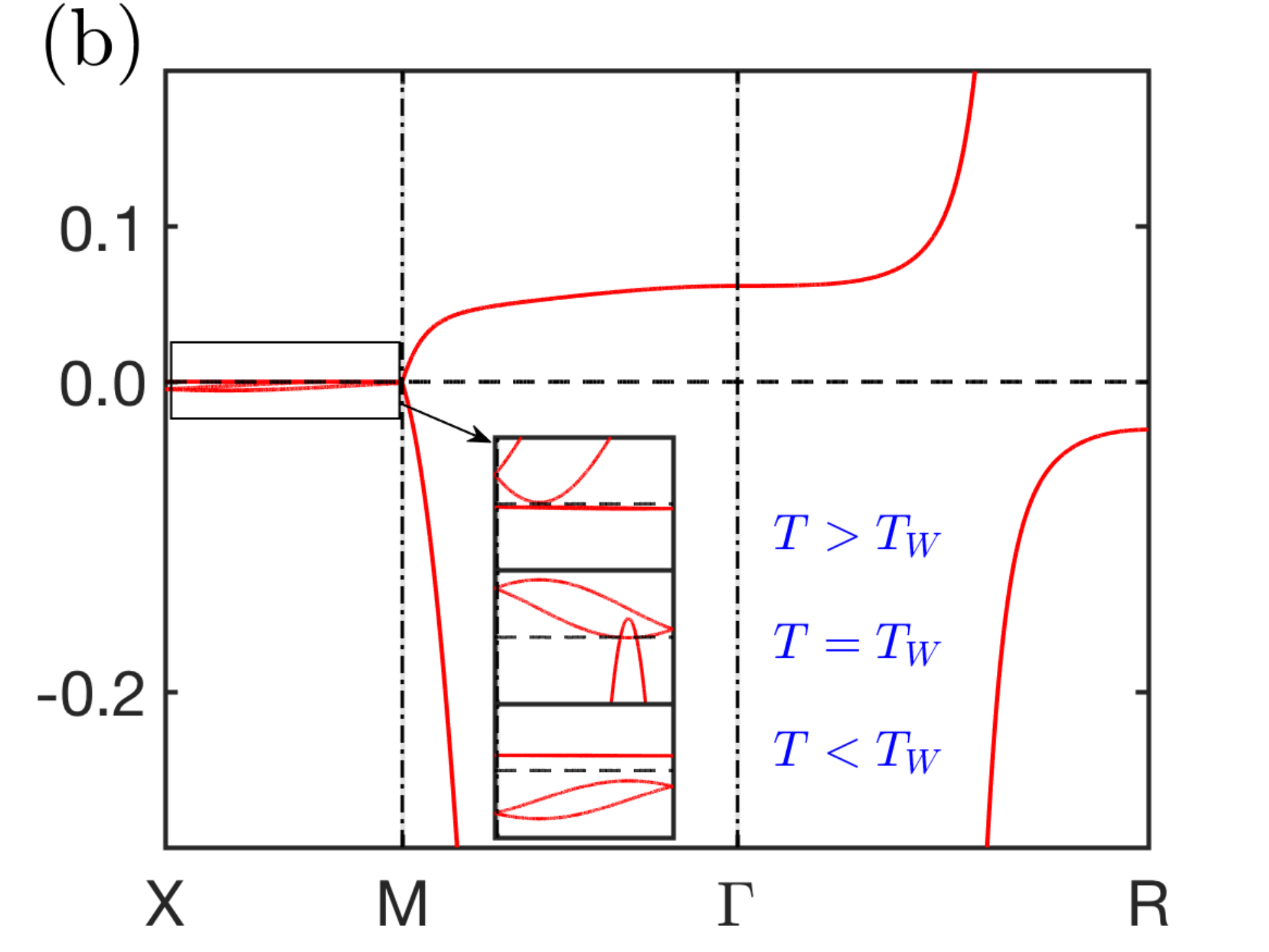}
\caption{(a) Topological phase diagram when the hybridization is governed by Rashba interaction. Here $z$ axis is the high symmetry axis and the anisotropy of hopping amplitude is characterized with $a_{x}=1$, $a_{y}=0.5$, $a_{z}=0.08$.  Gray, green, blue, and purple regimes are Weyl semi-metallic phases with Weyl nodes emerge at at $\mathbf{k}_{\text{w}}=(0,0,k_{\text{w}_{z}})$, $(0,\pi,k_{\text{w}_{z}})$,$(\pi,0,k_{\text{w}_{z}})$, and $(\pi,\pi,k_{\text{w}_{z}})$ respectively. (b)Emergence of the  Weyl semi-metallic phase at finite temperature. Here Weyl nodes emerge in $k_z$ axis from $X$ $(\pi,0,0)$ to $M$ $(\pi,0,\pi)$.   Note that two intersecting points in the inset may look like a nodal-ring. Clear demonstration of these intersecting points being Weyl points is confirmed by similar plots shown  in Fig.~\ref{fig6}. Parameters taken are  $t=1$, $\lambda_{Ra}=0.2$, $\eta=0.05$, $\varepsilon_{d}=0.978$, $\tilde{\lambda}_{so}=0.001$, $a_{y}=0.5$, and $a_{z}=0.001$. The critical temperature for emergence of Weyl nodes is $T_{W}=0.03$.}
\label{fig7}
\end{figure}

We now consider the Rashba spin-orbit hybridization interaction, $\mathbf{V}^2_{\mathbf{k}}$. In this case, as we shall show that instead of $\pm 2$, the monopole charge associated with Weyl node is  $\pm1$. In this case, 
 the Hamiltonian becomes
\begin{eqnarray}\label{HRa}
h_{\mathbf{k}}&=&-\mu_{\mathbf{k}}\tau_{0}\otimes \sigma_{0}+m_{\mathbf{k}} \tau_{z}\otimes \sigma_{0} +2r\lambda_{Ra} \tau_{x}\otimes  \boldsymbol{\sigma} \cdot \sin \mathbf{k}_{Ra} \\ \nonumber
&+&\tilde{\lambda}_{so}(\tau_{0}+\tau_{z})\otimes \boldsymbol{\sigma} \cdot \sin \mathbf{k}+\eta r^{2}\tilde{\lambda}_{so}(\tau_{z}-\tau_{0})\otimes \boldsymbol{\sigma} \cdot \sin \mathbf{k},
\end{eqnarray}
where we have denoted $\sin \mathbf{k}_{Ra}=\sin \mathbf{k}_{2D}\times \hat{z}$. Comparing to the Hamiltonian corresponding $\mathbf{V}^1_{\mathbf{k}}$, it is clear that $\lambda_{Ra}$ and $\mathbf{k}_{Ra}$ simply replace $\lambda_{so}$ and $\mathbf{k}_{2D}$. However,  because $\sin \mathbf{k}_{Ra}\cdot \sin \mathbf{k} =0$,  we find that
the analytic form of the energy spectrum to Eq.(\ref{HRa}) is given by
\begin{eqnarray}
E_{\mathbf{k}}^{(\alpha,\beta)} &=& -\mu_{\mathbf{k}} +\alpha\tilde{\lambda}_{so}(1+\eta r^{2})\sqrt{\sin^{2}\mathbf{k}_{3D}}+ \nonumber \\
&& \beta \sqrt{\left[m_{\mathbf{k}} + \alpha(1-\eta r^{2})\tilde{\lambda}_{so} \sqrt{\sin^{2}\mathbf{k}_{3D}}\right]^{2}+4r^2\lambda_{Ra}^{2}\sin^{2}\mathbf{k}_{Ra}},
\end{eqnarray}
where the anisotropy hopping strength has been considered.  Furthermore, we find that $E^{+,\beta}$ remains gapful.  Since $E_{\mathbf{k}}^{(-,+)}-E_{\mathbf{k}}^{(-,-)}=2\sqrt{\left[m_{\mathbf{k}} -(1-\eta r^{2})\tilde{\lambda}_{so} \sqrt{\sin^{2}\mathbf{k}_{3D}}\right]^{2}+4r^2\lambda_{Ra}^{2}\sin^{2}\mathbf{k}_{Ra}}$,  the requirements of $\sin^{2}\mathbf{k}_{Ra}=0$ and $m_{\mathbf{k}}=(1-\eta r^{2})\tilde{\lambda}_{so}\sqrt{\sin^{2}\mathbf{k}_{3D}}$ give rise to the condition for emergence of Weyl semi-metallic phase as 
\begin{equation} 
(\varepsilon_{\mathbf{k}_{\text{w}}}-\varepsilon_{\lambda})^{2}=4\tilde{\varepsilon}_{\tilde{\lambda}_{so}}^{2}\sin^{2}k_{\text{w}_{z}},
\end{equation}
where the effective parameter that determines the phase boundary is given by $\tilde{\varepsilon}_{\tilde{\lambda}_{so}} = (1-\eta r^{2})\tilde{\lambda}_{so}/(1+\eta r^{2})$. Similarly, solutions to Eq.(\ref{HRa}) form the boundary curve between the Weyl semi-metallic phase and insulating phases as illustrated in Fig.~\ref{fig7}(a). In addition, solving mean-field equations, Eqs.(\ref{meanM1}) and (\ref{meanM2}) enables one to find that it is possible to tune the Kondo insulator across Weyl semi-metallic phases at finite temperatures.  As illustrated in Fig. \ref{fig7}(b), the transition occurs at $T_W=0.03$ when parameters are taken at $t=1$, $\lambda_{Ra}=0.2$, $\eta=0.05$, $\varepsilon_{d}=0.978$, $\tilde{\lambda}_{so}=0.001$, $a_{y}=0.5$, and $a_{z}=0.001$.It is seen that the critical temperature for emergence of Weyl nodes is $T_{W}=0.03$. In addition, arranging the linearized form of Eq.(\ref{HRa}) in the form of Eq.(\ref{chargematrix}), we find 
\begin{equation}
\xi_{\mathbf{k_{\text{w}}}+\mathbf{q}}^{\pm} = \left[ -\mu_{\mathbf{k_{\text{w}}}+\mathbf{q}}\pm (1+\eta r^{2}) K\right] \tau_{0}+\left[ m_{\mathbf{k_{\text{w}}}+\mathbf{q}}\pm (1-\eta r^{2}) K\right] \tau_{z}+2ir\lambda_{Ra}q_{\pm}\tau_{+}+\text{H.c.}
\end{equation}
It is clear that $\xi_{\mathbf{k_{\text{w}}}+\mathbf{q}}^{+}$ is gapful, while $\xi_{\mathbf{k_{\text{w}}}+\mathbf{q}}^{-}$ can be tuned through the Weyl nodal point. The monopole charge of the emergent Weyl node, however, exhibit charge $\pm1$\cite{Bernevig}, in contrast to the double Weyl node for the case with the hybridization matrix $\mathbf{V}^1_{\mathbf{k}}$.

\section{Conclusion and discussion}

In conclusion, we have demonstrated that by including spin-orbit interactions, topological Weyl semi-metallic phases generally emerge from a Kondo insulator upon either change of temperature or spin-orbit interactions. Two different symmetry classes for the emergent  topological semi-metallic phases can be realized in the Anderson lattice: inversion symmetric semi-metallic phase and time reversal invariant semi-metallic phase.  For inversion symmetric semi-metallic phase, we find that Weyl nodes appear in pairs with opposite charges ($\pm1$) that are split off from a Dirac node upon time-reversal symmetry broken. On the other hand, we find that the Weyl nodal-ring semi-metallic phase generally emerges when the inversion symmetry is broken in the Anderson lattice with general hybridization between the conduction electrons and electrons in $f$ orbit. Furthermore, when the inversion symmetry is broken 
through the bulk spin-orbit interaction, two pairs of Weyl nodes emerge together. Depending on the nature of spin-orbit interaction in the hybridization, the emergent Weyl semi-metallic phase can host Weyl nodes with monopole charges being $\pm 1$ or double Weyl nodes with charges being $\pm 2$.
All of these topological semi-metallic phases are shown to be accessible by tuning either temperature or spin-orbit interactions at the integer filling of two\cite{filling}.  In addition, when the filling of electrons is tuned away from the integer filling for Dirac or Weyl semi-metallic phases, the system becomes doped topological semi-metals and is a hole Fermi liquid or an electron Fermi liquid depending on the filling\cite{Mou}. Furthermore, by tuning the filling, it is expected that the system can be driven through the quantum critical point between the hole Fermi liquid and the electron Fermi liquid or be driven into the Dirac liquid/Weyl liquid regime\cite{Sheehy, Mou}, controlled by the quantum critical point.

While so far in this work we only consider results based on the slave-boson mean-field theory, we expect that our results are  robust qualitatively in 
the presence of correlation effects as long as the symmetry of the system is not changed. In particular, following Ref. [\onlinecite{Mou}], the quasi-particle lifetime $\tau$ near the Weyl node can be estimated to be the order:
$\tau^{-1} \sim \left( \frac{rV_K}{\epsilon_d +\lambda - \mu} \right)^2 \frac{\omega^2 + \pi^2 (k_BT)^2}{2(\epsilon_d +\lambda - \mu)}$, where $\hbar \omega$ is the energy of the quasi-particle and
$V_K$ is the hybridization at the mean field Fermi momentum.
Substituting numerical values, we find that $\epsilon_d +\lambda - \mu \sim 0.02t \sim {\rm 1-10meV}$ and $rV_K \sim 0.002t$. As a result, for quasi-particles of typical energy scales up to $10$ meV, $\tau^{-1} \sim 0.1 $meV and the broadening effect is limited for $k_BT \le 10$meV. The emergent topological semi-metallic phases predicted in this work are thus well defined in finite temperatures up  $k_B T \sim 10$meV. Our results thus reveal the unusual interplay between the topology of the electronic structures and the Kondo screening in the strongly correlated Anderson lattices and pave a way for systematically engineering topological semimetals based on Kondo lattice systems.  

\begin{acknowledgments}
This work was supported by the Ministry of
Science and Technology (MoST), Taiwan. We also acknowledge support
from the Center for Quantum Technology within the
framework of the Higher Education
Sprout Project by the Ministry of Education (MOE) in Taiwan.
\end{acknowledgments}

\end{document}